# A generalizable 3D framework and model for self-supervised learning in medical imaging

Tony Xu[1], Sepehr Hosseini[2], Chris Anderson[3], Anthony Rinaldi[4], Rahul G. Krishnan[2,5,6], Anne L. Martel[1,4], Maged Goubran[1,4,7,8,†]

[1] Department of Medical Biophysics, University of Toronto, Toronto, Ontario, Canada
[2] Department of Computer Science, University of Toronto, Toronto, Ontario, Canada
[3] Institute for Aerospace Studies, University of Toronto, Toronto, Ontario, Canada
[4] Physical Sciences Platform, Sunnybrook Research Institute, Toronto, Ontario, Canada
[5] Vector Institute, Toronto, Ontario, Canada
[6] Department of Laboratory Medicine and Pathobiology, University of Toronto, Toronto, Ontario, Canada
[7] Hurvitz Brain Sciences, Sunnybrook Health Sciences Centre, Toronto, Ontario, Canada
[8] Harquail Centre for Neuromodulation, Sunnybrook Health Sciences Centre, Toronto, Ontario, Canada

[†]Corresponding author:
maged.goubran@utoronto.ca

## Abstract

Current self-supervised learning methods for 3D medical imaging rely on simple pretext formulations and organ- or modality-specific datasets, limiting their generalizability and scalability. We present 3DINO, a cutting-edge SSL method adapted to 3D datasets, and use it to pretrain 3DINO-ViT: a general-purpose medical imaging model, on an exceptionally large, multimodal, and multi-organ dataset of ~100,000 3D medical imaging scans from over 10 organs. We validate 3DINO-ViT using extensive experiments on numerous medical imaging segmentation and classification tasks. Our results demonstrate that 3DINO-ViT generalizes across modalities and organs, including out-of-distribution tasks and datasets, outperforming state-of-the-art methods on the majority of evaluation metrics and labeled dataset sizes. Our 3DINO framework and 3DINO-ViT will be made available to enable research on 3D foundation models or further finetuning for a wide range of medical imaging applications.

## Introduction

Deep learning (DL) methods can enhance existing workflows for a variety of clinical tasks involving medical images [1–3], including detection [4–7], diagnosis [8–13], and risk profiling [14–18]. However, the data-hungry nature of these methods poses practical challenges for training and generalizability. Creating detailed labels for training DL models for *3D* medical imaging modalities is particularly time-consuming and expensive. To alleviate this, self-supervised learning (SSL) approaches have been proposed to reduce reliance on detailed ground truth annotations by leveraging unlabelled datasets [19–21]. Yet, most existing approaches for 3D medical imaging modalities train SSL methods using simple pretext formulations on unlabeled datasets that are similar to their downstream applications. Employing SSL-pretrained models on downstream datasets from similar modalities, organs, image characteristics, and distributions limits their generalizability and scalability. Notably, this single-distribution approach results in additional training overhead, as separate models would need to be pretrained for each downstream task for optimal performance. This can be further exacerbated by several factors, including the rarity of the disease, the ability to acquire high-resolution, multidimensional data, or the scarcity and cost of certain imaging modalities. The availability of *general-purpose* pretrained weights could facilitate more widespread adoption of DL in medical imaging applications by greatly boosting the accuracy of DL models in these label-scarce regimes.

Recent work in SSL has scaled to larger and highly diverse pretraining datasets, with models able to create image representations that are generalizable to many downstream tasks [22–26]. While these pipelines are capable of achieving state-of-the-art (SOTA) results for *2D benchmarks*, scaling them to 3D data is computationally prohibitive, requiring large datasets, batch sizes (often ranging from 512-4096), and long train times to learn effectively. One way to resolve these constraints is to cast the 3D SSL task as a 2D task by viewing 3D images slice-by-slice. However, previous studies have shown that keeping the full 3D anatomical context is important when applying DL to medical images and clinical scenarios [19,27]. The recently proposed DINOv2 [28] SSL pipeline provides numerous improvements in accuracy and computational efficiency relative to its counterparts, making it a strong candidate for generating genuinely *3D* image representations.

In this work, we develop *3D-DINOv2 (3DINO):* a cutting-edge and memory-efficient framework adapting DINOv2 to 3D medical imaging inputs, and present the 3DINO-Vision Transformer (3DINO-ViT): a general-purpose ViT [29] model pretrained on an exceptionally large, multimodal, and multi-organ dataset of nearly 100,000 unlabeled 3D medical volumes curated from 35 publicly available and internal data studies (**Fig. 1**). We specifically acquired datasets consisting of MRI (N = 70,434) and CT (N = 27,815) volumes, with a small brain PET (N = 566) dataset (**Fig. 1**b). The original DINO [24] is formulated as a form of self-distillation between a student network and an exponential moving average (EMA) teacher network. Building on this, DINOv2 introduces a host of efficiency improvements, objectives, and regularization methods that stabilize training and reduce training time by 2x and GPU memory usage by 3x. The pretext formulation combines an image-level objective and a patch-level objective, where original volumes are augmented to generate two global and eight local crops (a total of 10 augmentations for the objectives per scan). 3DINO augmentations must act on 3D grayscale images, hence we use a custom implementation of RandomResizedCrop, and other augmentations such as random contrast adjustment, additive noise, and gibbs noise. To create 3DINO, we convert the image-specific operations from 2D to 3D, while making minimal changes to the ViT layers. By doing so, we ensure that we take advantage of all high- and low-level efficiency improvements introduced in the original work. We additionally modify the 3DINO-ViT model backbone to enhance its performance on downstream segmentation tasks by converting an adapter module to 3D inputs (*3D ViT-Adapter*). This module has been employed in 2D images to inject spatial inductive biases into pretrained ViT models for dense (pixel-level) tasks [30]. To achieve this, we implement a 3D version of multi-scale deformable attention (MSDA) [31]. To our knowledge, while large SSL-pretrained models have been proposed for a single 3D medical imaging modality [21], we introduce the first 3D SSL-based medical imaging model that can extract salient features and generalize across multiple modalities simultaneously. The full 3DINO pipeline, 3D ViT-Adapter, and 3D MSDA, along with our 3DINO-ViT weights, are made available at https://github.com/AICONSlab/3DINO to facilitate research toward 3D medical imaging foundation models or further finetuning over a vast range of medical applications across numerous organs and modalities.

# Results

We compare the efficacy of 3DINO-ViT weights on downstream tasks against four other pretraining and initialization methods. The first comparison randomly initializes the ViT network and trains it end-to-end from scratch (‘Random’). As a SOTA *pretrained* medical imaging backbone, we utilize the Sliding Window (Swin) ViT from Tang et al. [21] (‘Swin Transfer’). To account for differences between model architectures (3DINO-ViT uses a vanilla ViT and Swin Transfer uses a Swin ViT), we also employ a randomly initialized Swin ViT to evaluate the *relative* benefits of using pretrained weights (‘Swin Random’). Finally, we perform a more direct comparison against the 3DINO pretraining method by adapting the SOTA Swin ViT pretraining framework from Tang et al. [21] for a vanilla ViT (‘MONAI-ViT’), and pretrain it on the same unlabeled dataset.

## Within and out-of-distribution downstream tasks

To evaluate the saliency and generalizability of 3DINO-VIT pretrained weights, we use popular medical image segmentation and classification benchmarks/challenges (Methods – Finetuning methods). As a segmentation benchmark, we use the 2021 Brain Tumor Segmentation (BraTS) Challenge [32] for MRI, which looks to delineate between Tumor Core, Enhancing Tumor, and Whole Tumor regions in glioblastoma patients. We additionally use the Beyond the Cranial Vault (BTCV) Challenge [33] for a 14-class CT abdominal organ segmentation task. We further evaluate the generalizability of 3DINO pretraining on unseen (out-of-distribution) organs and 3D modalities with minimal presence in the pretraining dataset. Specifically, we evaluate 3DINO’s performance on left atrium MRI segmentation (LA-SEG) [34], an unseen organ, and 3D breast ultrasound tumor segmentation (TDSC-ABUS) [35], a completely unseen modality. For classification tasks, we investigate brain age classification on the MRI ICBM dataset [36] and use the COVID-CT-MD lung CT dataset [37] to classify between healthy patients, those with community-acquired pneumonia (CAP), and individuals with Novel Coronavirus (COVID-19). We further perform experiments using different amounts of labeled training data by randomly subsampling a certain percentage of the full labeled dataset.

## Performance on segmentation and classification tasks

For the Random, Swin Random, and Swin Transfer networks, segmentation was performed via appending convolutional decoder heads to pretrained encoders and tuning end-to-end. 3DINO-ViT and MONAI-ViT weights are *frozen*, and a 3D-ViT Adapter module and simple convolutional decoder are tuned (**Extended data Fig. 1**). 3DINO yielded significantly improved segmentation results relative to all SOTA techniques on all evaluation metrics in most comparisons ($p < 0.05$; **Fig. 2**a**, 2**b**, 2**e). 3DINO-ViT was able to jointly improve representations for both segmentation tasks in all percentages of labeled data, including when using the full labeled dataset. The pretrained weights significantly improved performance at all percentages of labeled data relative to the Random encoder (BraTS with 10% data: 0.90 (0.88, 0.91) Dice for *3DINO-ViT* vs 0.87 (0.85, 0.89) for *Random*; BTCV with 25%: 0.77 (0.72, 0.81) *3DINO-ViT* vs 0.59 (0.53, 0.65) for *Random*). For both segmentation tasks, 3DINO-ViT trained using less than 50% of all labeled data achieved statistically (**Fig. 2**) and visually (**Fig. 3-4, Extended data Fig. 2-6**) comparable results to other baselines trained using 100% of labeled data. However, when using 100% of the labeled dataset, the relative improvements of 3DINO-ViT over the next best baseline are reduced and were not significant, with 0.8% Dice improvement in BraTS and 0.9% in BTCV. Results across other evaluation metrics are presented in the Supplementary Material, highlighting the same trend of 3DINO-ViT's improved segmentation results over other SOTA pretrained models.

For classification tasks, we trained a linear classifier on top of all pretrained networks without finetuning the pretrained weights. We use MONAI-ViT as one comparison and take the pretrained contrastive head from the Swin Transfer network as another. Despite the tasks' difficulty and sparser ground truth, the proposed 3DINO-ViT performs universally better than other models ($p < 0.05$; **Fig. 2c-d**). Averaging over all dataset sizes, 3DINO-ViT obtained an 18.9% higher area under the receiver operating characteristic curve (AUC) on COVID-CT-MD, with a particularly notable increase of 23% AUC on classifying patients with COVID-19 relative to the next best baseline. On ICBM, an average of 5.3% higher AUC was obtained with a 22% AUC improvement for classifying individuals aged [40, 50) years over the next best baseline (**Fig. 2e-f**, example cases in **Fig. 5**). These experiments, conducted with completely frozen pretraining weights, further highlight the saliency of the learned representations for different downstream tasks.

## 3DINO patch-level representations for image segmentation

As a standalone comparison of the effectiveness of 3DINO patch-level representations for image segmentation, we performed experiments with a lightweight segmentation decoder. We froze 3DINO-ViT's pretrained weights and finetuned a two-layer *linear network* on downstream tasks (Methods – Linear decoder segmentation). We compared the performance against both the Random and MONAI-ViT networks (**Fig. 2**g), and found 3DINO-ViT achieved a significant improvement of 61% Dice over the next best baseline on BTCV ($p < 0.05$), and 15% on BraTS ($p < 0.001$).

## Performance on unseen organs and modalities

On the out-of-distribution tasks, 3DINO-ViT significantly outperformed other SOTA methods, with 1.9% improved Dice on left atrium segmentation, and 24% in 3D ultrasound tumor segmentation over the next best baseline, when finetuning with 25% of the labeled dataset (**Fig. 6**). Though this improvement drops to 1.1% and 0.8% respectively, 3DINO-ViT maintains its advantage over other baselines even when tuning with 100% of the labeled dataset. This demonstrates the capability of 3DINO to create *generalizable* weights that can be applied to image distributions unseen during pretraining.

# Discussion

In this work, we introduced 3DINO, a cutting-edge, computationally efficient 3D SSL framework, by modifying the highly scalable DINOv2 SSL pretraining methodology to perform representation learning on 3D medical images. We applied 3DINO pretraining to a curated, ultra-large, multi-organ, and multimodal medical imaging dataset to build a general-purpose pretrained ViT model (3DINO-ViT). To enhance 3DINO pretraining when applied to dense tasks, we created and used a 3D ViT-Adapter when evaluating on segmentation baselines. Altogether, we observed that the full 3DINO pipeline improved the ViT's data efficiency and generalizability on both pixel-level (segmentation) and image-level (classification) tasks. Transfer learning with frozen 3DINO-ViT weights improved performance over other SOTA methods on the majority of evaluation metrics at all dataset sizes and for all tasks, including when using the full labeled dataset. Finally, we demonstrated 3DINO-ViT's capability to generalize to out-of-distribution data from unseen organs and modalities in pretraining.

On the primary Dice similarity segmentation metric, 3DINO-ViT consistently outperformed the compared pretrained models. When comparing the amount of improvement with 3DINO-ViT initialization over Random initialization relative to Swin Transfer initialization over Swin Random initialization, we noted that 3DINO pretraining yielded relatively larger benefits to segmentation performance. The maximum Dice improvement for 3DINO-ViT over the Random encoder was 13.0% on BraTS and 55.1% on BTCV, whereas the Swin Transfer network versus Swin Random improved 5.1% on BraTS and 1.8% on BTCV, respectively (**Fig. 2)**. 3DINO-ViT and MONAI-ViT have the same network architecture, segmentation setup, and pretraining dataset, differing only in pretraining methodology. While MONAI-ViT generally improved over Random initialization, we found 3DINO pretraining to generate more salient representations for both segmentation and classification tasks. We expect this can be attributed jointly to the quality of the pretext task and the computational efficiency of 3DINO, as we were only able to use a batch size of 512 when pretraining 3DINO-ViT, compared to 64 for MONAI-ViT. In line with other works in SSL, we found the effect of pretraining was more pronounced when using less data for finetuning. Finally, by greatly simplifying the segmentation setup to train solely a linear network to act on patch-level representations, we showed that 3DINO-ViT features are more directly suitable to segmentation tasks than those output by other ViT initialization methods (**Fig. 2**g).

We similarly found 3DINO-ViT outperformed all other pretrained models in classification tasks in all metrics. Classification with 3DINO-ViT also yielded improved class-wise pseudo-probabilities. For instance, in particular cases on brain age classification on the ICBM dataset where individuals were in between two classification bins, we observed 3DINO was able to successfully classify them and better capture the uncertainty in the prediction (**Fig. 5**a). We note that the Swin Transformer pretraining method introduced by Tang et al. [21]

was not originally intended to generate image-level representations for classification tasks. Thus, to the best of our knowledge, and based on our extensive validation, 3DINO is the first pretraining method for 3D medical images that is able to perform both image-level and patch-level tasks, while jointly improving performance on both.

We probed the out-of-domain generalizability of 3DINO using heart MRI and breast US segmentation. 3DINO significantly improved over other SOTA methods in both of these tasks, demonstrating its ability to learn features that are salient even for unseen distributions. Perhaps owing to the lower difficulty of the task, we found the benefit of 3DINO in left-atrium segmentation to be relatively low when finetuning with 25% of the labeled dataset (1.9% improvement), though this improves under smaller dataset sizes (9.3% improvement at 10% dataset size, and 23% improvement at 5%).

To visually investigate the saliency of 3DINO-ViT representations, we generated principal component analysis (PCA) and multi-head self-attention (MHSA)-based visualizations of the representations (**Fig. 1**c**, 6**d**, 6**h). PCA visualizations demonstrate that common modes of variation for all datasets are between background versus foreground, outlining the surface of the organ, and varying across anatomical axes. Principal components generated on BraTS images found inside the tumor extent were often distinct from other brain tissues. We note that visualizations on the completely unseen US modality (**Fig. 6**h) were relatively less salient than those generated in previously seen modalities.

We also consider 3DINO-ViT relative to recently proposed “foundation models” in 3D medical image segmentation [38], which build upon Segment Anything Models (SAM) [39]. Since SAM-like networks are trained using labeled data and 3DINO is a self-supervised pretraining method, both methods are synergistic. The original SAM paper used weights from SSL pretraining to initialize the image encoder network [39]. 3DINO thus represents a novel ViT pretraining method for 3D inputs that is able to act as an initialization step for 3D SAM or other methods.

One limitation of this study is that it primarily uses MRI and CT data for pretraining and the dataset does not encompass a balanced list of organs. Despite this, the out-of-domain generalizability of the method was explored on cardiac MRI and breast ultrasound images, and found to be significantly better than other techniques (**Fig. 6**). While 3DINO-ViT is highly efficient in the number of trainable parameters because it is frozen during finetuning, the full segmentation pipeline is relatively high in runtime complexity. While we were not able to perform these experiments due to computational limitations, we expect using the largest ViT network (ViT-Giant) and equivalent computational resources (8x larger GPU memory than what we used) to the original DINOv2 to improve overall performance, along with better hyperparameter tuning.

The formulation of 3DINO addresses key prior limitations of SSL pipelines for 3D medical imaging in terms of generalizability and computational complexity, leading to promising results in downstream tasks and intuitive unsupervised representations. By leveraging 3DINO-ViT, we can reduce the amount of labeled data needed for diverse downstream medical imaging tasks without requiring expensive model retraining on in-domain unlabeled datasets, enabling generalizable and data-efficient models. The presented pipeline and models could be highly beneficial when finetuned across a wide variety of challenging applications and tasks in medical imaging, especially in environments with limited access to detailed annotations and resources.

# Figures

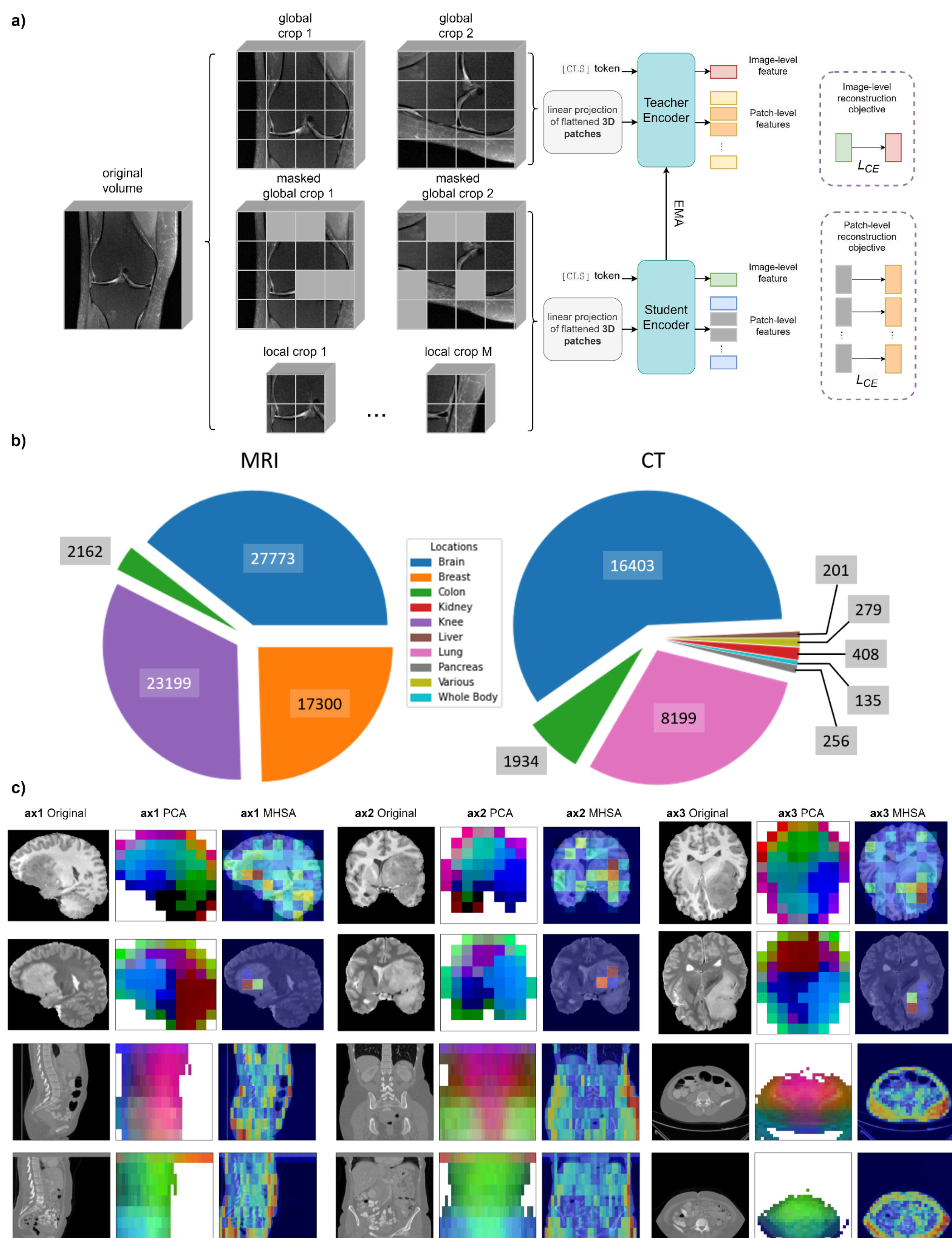

***Figure 1**: Overview of 3DINO methodology and large pretraining dataset.* ***a)*** *3DINO combines an image-level objective and a patch-level objective. Original volumes are randomly augmented twice to create global crops, and augmented eight times to yield local crops. The image-level objective is taken by distilling the* [CLS] *token representations between the student and* exponential moving average (EMA) *teacher networks. The patch-level objective is computed between patch representations at masked regions in the student network input and corresponding unmasked EMA teacher representations.* $L_{CE}$ *indicates Cross-Entropy loss, with the final 3DINO loss consisting of the summed image-level distillation and patch-level reconstruction objectives.* ***b)*** *Breakdown of large*

*multimodal, multi-organ pretraining dataset of 100,000 3D scans with over 10 organs from 35 publicly available and internal studies (the number of volumes per modality per anatomical location/organ; MRI = 70,434 volumes, CT = 27,815, and PET = 566)* ***c)*** *Original image,* principal component analysis (PCA) *on patch-level representations,* and multi-head self-attention (MHSA) *attention map visualized for three image planes. Each row in order: BraTS T1-weighted, T2-weighted, and two patients from BTCV. PCA visualizations are obtained per image from patch-level representation vectors. The first (in terms of explained variance) PCA component is used to mask image background (white) with a simple threshold, with the next three normalized and mapped to RGB channels.* MHSA *attention maps are obtained from the* [CLS] *token of final 3DINO-ViT layer. Images were not registered to atlases for visualization or training/testing.*

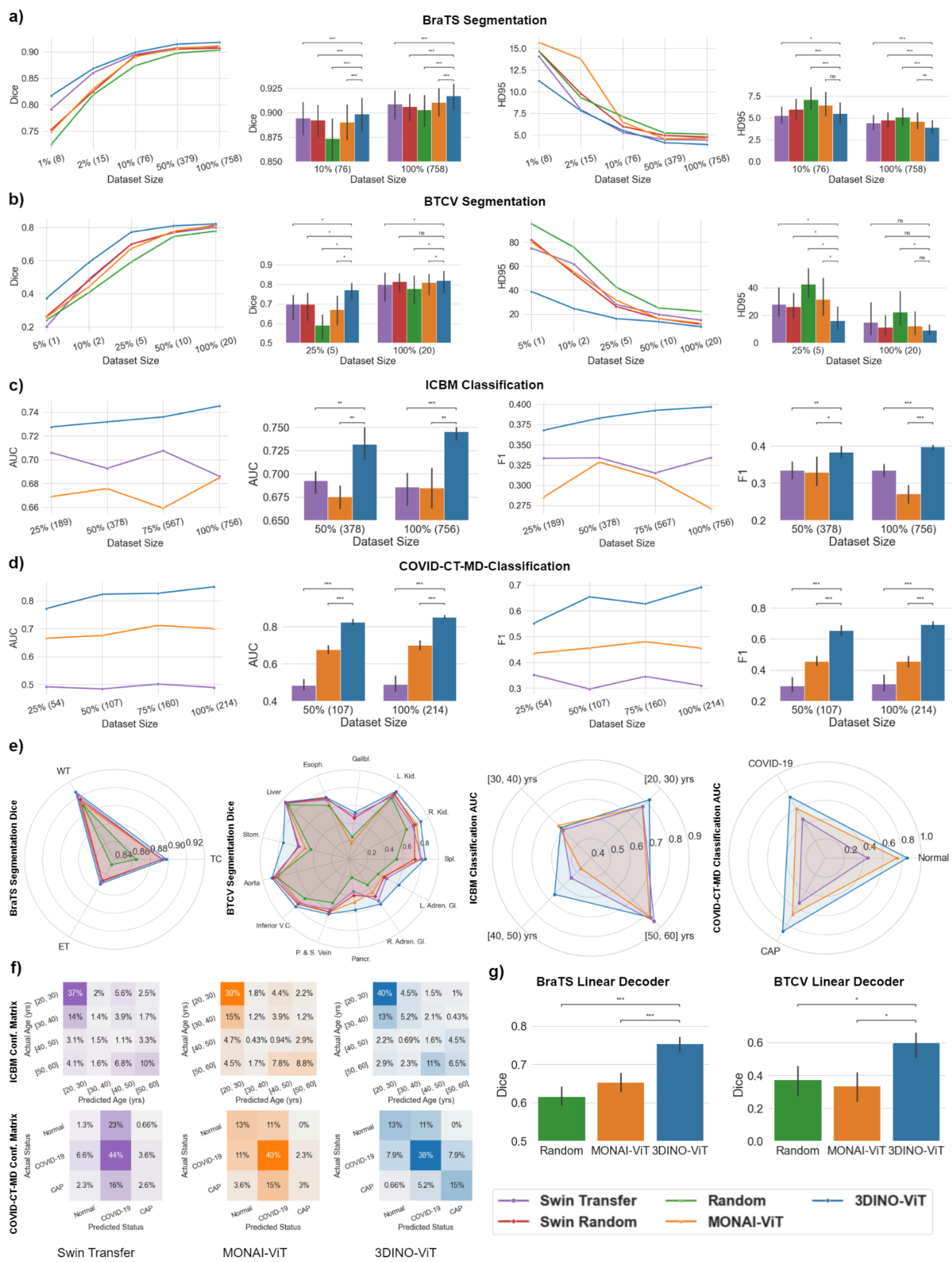

***Figure 2**: Evaluation and comparison to SOTA pretrained models on the BraTS and BTCV segmentation, and ICBM and COVID-CT-MD classification tasks. **a)** BraTS segmentation Dice scores and 95th percentile Hausdorff Distance (HD95). **b)** BTCV segmentation Dice scores and HD95. **c)** ICBM classification AUC and F1 scores. **d)** COVID-CT-MD classification AUC and F1 scores. **a)-d)** Finetuning results with multiple sizes of labeled dataset, x-axis displays training dataset size in a percentage of the full dataset, with actual number of labeled samples in parentheses. Bar plots compare the third-largest and largest training dataset sizes with error bars in **a)**, **b)** from the 95% bootstrapped confidence interval (CI) of item-wise metrics and error bars. Bar plots in **c)**, **d)** are*

*obtained from 95% bootstrapped CI of metrics obtained from five separate experiments with randomly subsampled training sets (100% data comparison is equivalent to adjusting experiment seed). Statistical significance in **a)**, **b)** is computed via a paired nonparametric Wilcoxon test on metrics per item. Significance in **c)**, **d)** are computed via an unpaired Welch's t-test against metrics per experiment. **e)** Plots comparing per-class Dice/AUC scores for segmentation/classification experiments using the third-largest training dataset size. **f)** Normalized and averaged classification confusion matrices using the third-largest training dataset size. **g)** Dice scores for linear decoder segmentation experiments.*

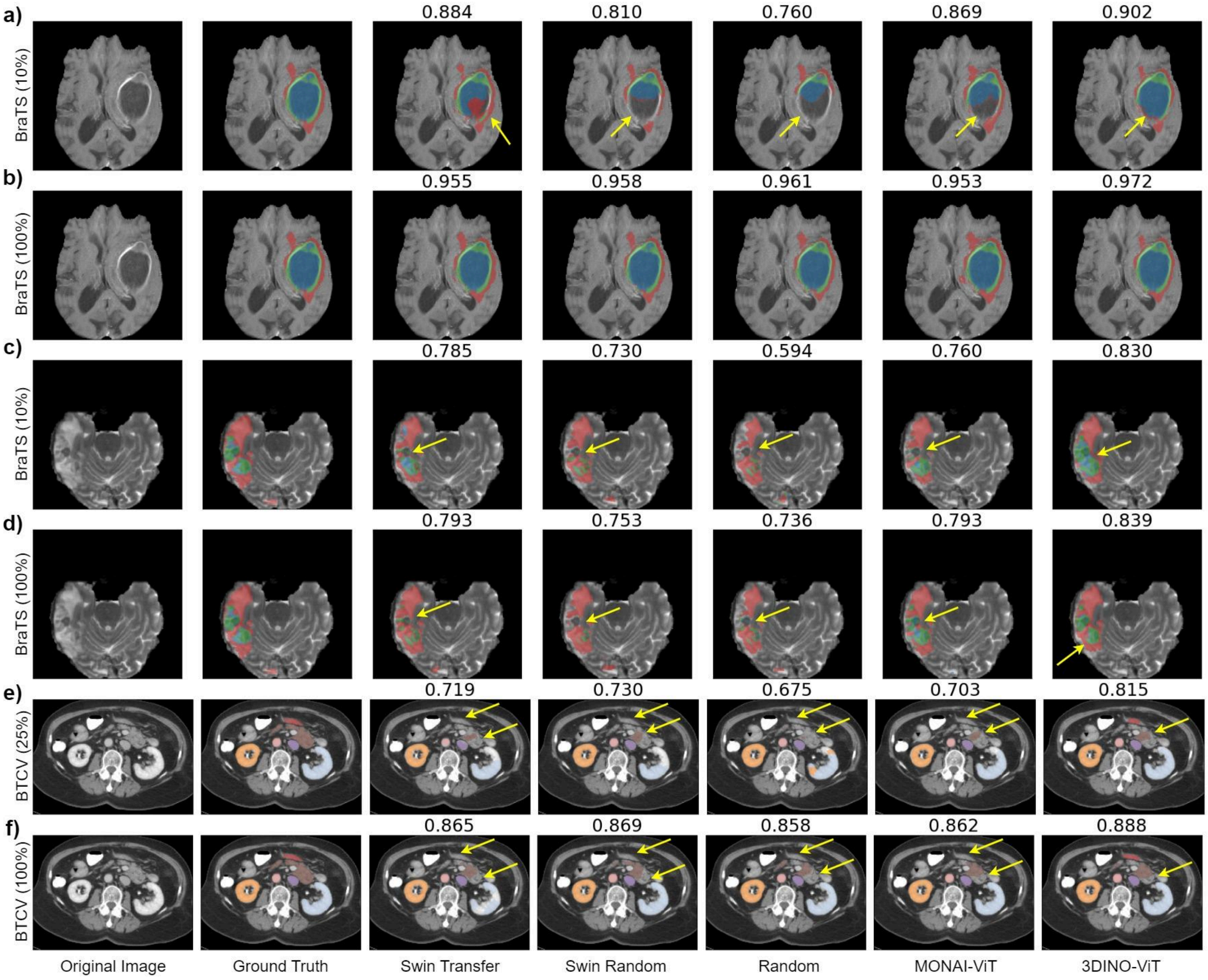

***Figure 3**: Visualization of segmentation predictions on BraTS and BTCV datasets. Original image, ground truth segmentation, and visualized segmentation outputs per pretraining methodology using the third-largest and largest training datasets. **a)**-**d)** BraTS labels are necrotic and non-enhancing tumor core (blue), peritumoral edema (red), and enhancing tumor (green). **e)**, **f)** BTCV organs visualized are: right kidney (light blue), left kidney (orange), stomach (red), aorta (pink), inferior vena cava (purple), and pancreas (brown). The numbers above images are Dice segmentation scores obtained on the full 3D volume. Arrows indicate degraded segmentation outputs in methods relative to ground truth segmentation. **a)** Contrast-enhanced T1-weighted image from BraTS with segmentation models trained using 10% of the full dataset size, and **b)** using 100% of the full dataset size. **c)** T2-weighted image from BraTS with segmentation models trained using 10% of the full dataset size, and **d)** using 100% of full dataset size. **e)** CT image from BTCV with segmentation models trained using 25% of the full dataset size, and **f)** using 100% of the full dataset size.*

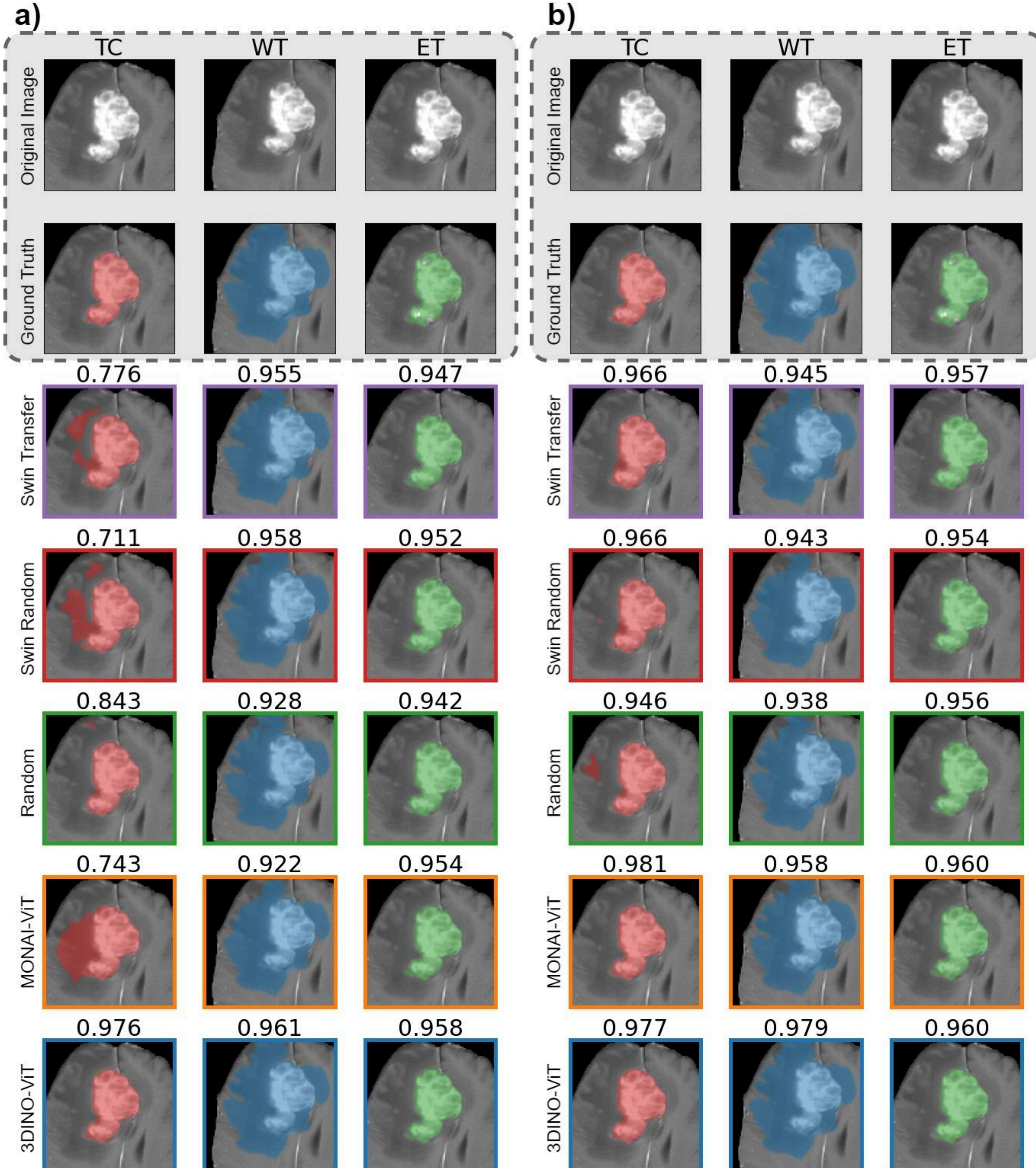

***Figure 4:*** *Class-separated visualizations of segmentation network predictions on BraTS. Above each image is the Dice score for segmentation per class. Visualized for* ***a)*** *models trained using 10% of the labeled finetuning dataset, and* ***b)*** *100% of the labeled finetuning dataset. TC: Tumor Core; WT: Whole Tumor; ET: Enhancing Tumor.*

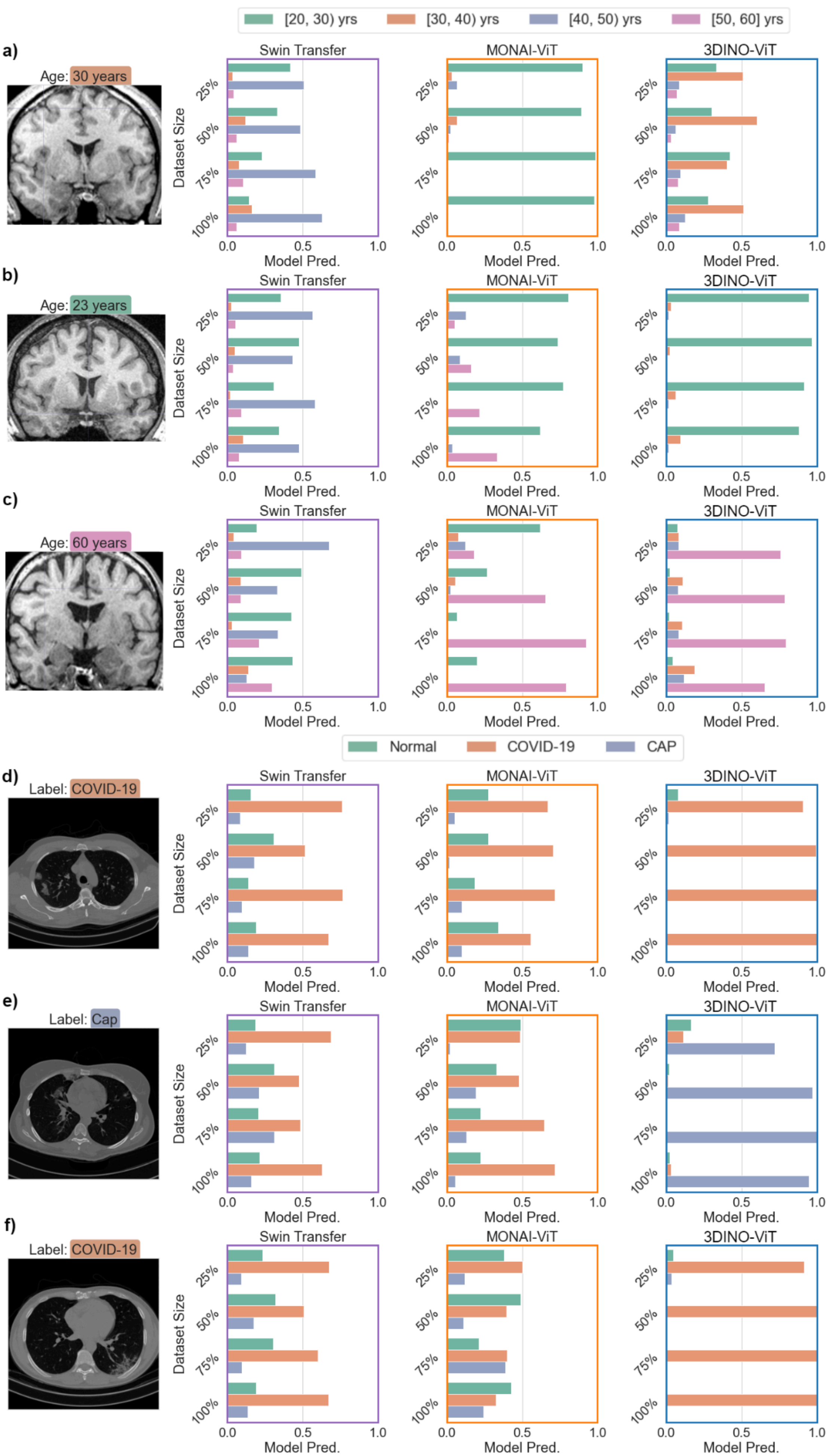

***Figure 5:*** *Visualization of classification performance on ICBM age prediction (**a)-c)**) and COVID-CT-MD disease prediction (**d)-f)**)for all methods and all classes at 25%, 50%, 75%, and 100% of finetuning data. **a)-f)** are separate cases in the testing dataset. Bar plot depicts the predicted class probabilities output by each model. Ground truth age is depicted above the image (color-coded). Pred: Prediction. Cap: community-acquired pneumonia.*

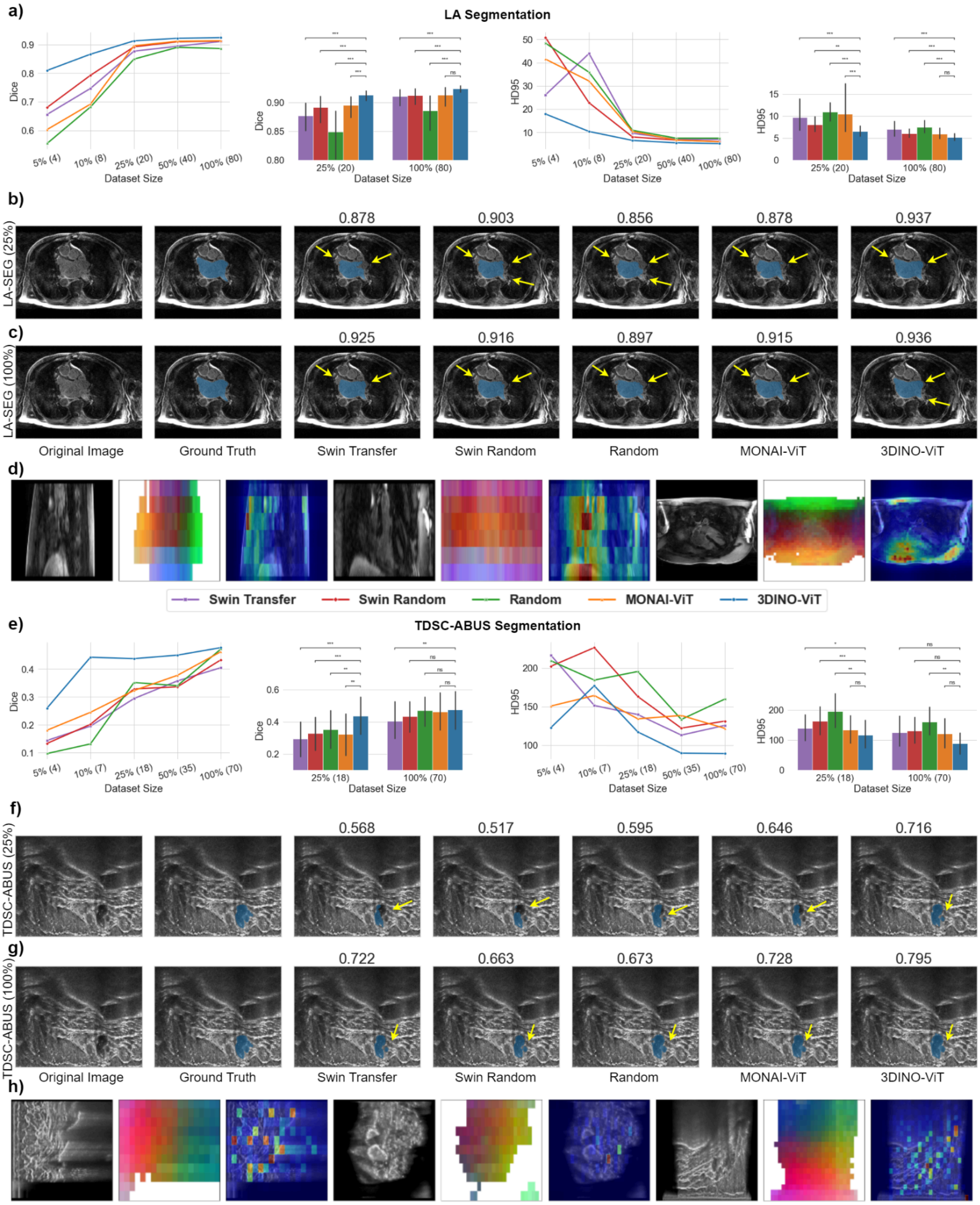

***Figure 6****: Model evaluation on out-of-distribution tasks: left atrium segmentation (LA-SEG; unseen organ) and 3D breast ultrasound tumor (TDSC-ABUS; unseen modality).* ***a), e)*** *Dice and HD95 scores for the LA-SEG and TDSC-ABUS segmentation tasks, respectively.* ***b), c), f), g)*** *Original image,*

*ground truth segmentation, and visualized segmentation per pretraining methodology using third-largest and largest finetuning dataset subsets. Yellow arrows indicate degraded model outputs relative to ground truth segmentations. The numbers above images are Dice segmentation scores obtained on the full 3D volume.* ***b)*** *LA-SEG visualization when finetuning using 25% of the full labeled dataset.* ***c)*** *LA-SEG visualization when finetuning using 100% of the full labeled dataset.* ***f)*** *TDSC-ABUS visualization when finetuning using 25% of the full labeled dataset.* ***g)*** *TDSC-ABUS visualization when finetuning using 100% of the full labeled dataset.* ***d)****,* ***h)*** *Unsupervised visualizations on random volumes sampled from the LA-SEG and TDSC-ABUS datasets respectively. Ordering and visualization methods are analogous to* ***Fig. 1c****.*

# Methods

## Pretraining Methods

### Pretraining unlabeled dataset

We constructed our multimodal 3D medical imaging pretraining dataset from a variety of publicly available datasets and one internal dataset shown in **Extended data Table 1**. The Sunnybrook Health Sciences Centre provided research ethics approval for the internal Acute Stroke dataset (REB #2430). We filtered pretraining datasets for excessively few DICOM slices (>24 slices) to avoid overly pixelated volumes in the cross-slice dimension (lower z-axis resolution). To reduce redundancy and perform a naive form of deduplication, we took random subsets of a few exceptionally large datasets. Subsets were taken from the FastMRI Knee dataset [40,41] and the RSNA Intracranial Hemorrhage Detection [42] dataset by randomly sampling half of the dataset. A subset of NLST [43,44] was taken by randomly sampling 500 patients, and a subset of 4D-Lung [45,46] was taken by sampling 30-40 volumes per patient. After deduplication and filtering for slice counts, we created a 3D medical imaging dataset of 98,815 unlabeled volumes. The high-resolution adaptation (Methods - DINOv2) dataset was created by filtering for >48 DICOM slices, which resulted in a higher resolution 53,758 volume subset of the original data.

### Image-level objective

The original DINO [24] method is a self-supervised self-distillation method consisting of a teacher, $g_{\theta_t}$ and student network, $g_{\theta_s}$ parameterized by $\theta_t$ and $\theta_s$ respectively. From an unlabeled medical image sampled from the dataset, $x$ , two randomly augmented "global crops", $x_1^g$ and $x_2^g$ , as well as $L$ randomly augmented "local crops", $\{x_1^l, x_2^l, ..., x_L^l\}$ are generated to create a set of crops $C$ . The global crops are passed through the teacher network, and all crops are passed through the student network. Given a global crop passed through the teacher, $x_1 \in \{x_1^g, x_2^g\}$ , the overall task of the student network is to predict the teacher representation using *all* other crops, $x_2 \in C, x_2 \neq x_1$ .

The feature representation output from the student and teacher are converted into probability distributions via a softmax function, $\sigma(\cdot)$ , and sharpened via a temperature parameter, $\tau$ . The student probability distribution is defined as $P_s(x) = \sigma(g_{\theta_s}(x)/\tau_s)$ , with the same formula using $P_t$ and $\tau_t$ for the teacher network.

The following image-level loss function summarizes the objective of DINO:

$$\mathcal{L}_{image} = \sum_{x_1 \in \{x_1^g, x_2^g\}} \sum_{x_2 \in C x_2 \neq x_1} H(P_t(x_1)^{[\text{CLS}]}, P_s(x_2)^{[\text{CLS}]})$$

Where $H(\cdot)$ is the cross-entropy function. Rather than explicitly training the teacher in this framework, it is obtained as an exponential moving average (EMA) of the student model: $\theta_t \leftarrow \lambda\theta_t + (1-\lambda)\theta_s$ . The original DINO method learns image-level representations by taking features learned from the classification [26] token of a ViT network, which is represented by the $^{[\mathrm{CLS}]}$ superscript. Hence, this task forms the "image-level objective".

## DINOv2

DINOv2 [28] makes several key improvements beyond the original DINO method that make it more scalable and efficient for learning representations from large (medical) images. Firstly, it introduces a masked image modeling (MIM) objective originally from the iBOT work [26]. This method masks patch regions for global crops passed to the student using a binary mask, $\mathbf{m} \in \{0,1\}^N$ over the $N$ patches comprising the crop. Using the masked crop, $\hat{x}^g$, the student model is tasked with predicting the teacher representations at the masked regions, leading to the loss function:

$$\mathcal{L}_{patch} = \sum_{i=1}^{N} \quad m_i \cdot H(P_t(x_1^g)^{[\mathtt{i}]}, P_s(\hat{x}_1^g)^{[\mathtt{i}]}) + H(P_t(x_2^g)^{[\mathtt{i}]}, P_s(\hat{x}_2^g)^{[\mathtt{i}]})$$

Where $m_i$ is the binary mask value, and $P(\cdot)^{[\mathtt{i}]}$ is the patch token output by the encoding model, both enumerated by patch location, $i$. By introducing this objective, the iBOT paper improved patch-level representation quality and robustness to image corruption [26]. This is key for dense downstream tasks like segmentation and improving representations in the presence of out-of-sample distribution shifts and corruptions/artifacts, which are abundant in medical imaging due to differences in scanners, imaging hardware, acquisition parameters and sequence design [47]. The final loss function adds patch-level loss to the image-level loss.

Additionally, DINOv2 introduces several improvements on computational and memory efficiency that enable larger batch training. These include improving the efficiency of computing self-attention, allowing nested tensors in self-attention, saving memory in the stochastic depth operation, and taking advantage of the new Fully-Sharded Data Parallel modules in PyTorch. Relative to iBOT, their code runs approximately 2 times faster with 1/3 of the GPU memory usage [28]. Finally, they introduce several regularization methods including Sinkhorn-Knopp centering [48] and the KoLeo regularizer [49] that stabilize training progress at scale and reduce the likelihood of model collapse.

DINOv2 also introduces a secondary high-resolution adaptation stage to the pretraining process. In segmentation tasks, maintaining image resolution is important for extracting smaller objects and features. However, training self-supervised methods from scratch with high-resolution inputs is highly computationally expensive. Instead, the authors found that introducing a short adaptation period using high-resolution inputs at the end of pretraining yields comparable results compared to full training using high-resolution [28]. We similarly introduce this adaptation stage into our method.

## 3DINO

To adapt the DINOv2 model architecture for 3D inputs, we adjusted the ViT encoder network to flatten and project 3D input patches. To permit variable-sized inputs, we implemented 3D interpolation of the learned position encoding vectors. To simplify the 3D masking operation, we randomly sample masking locations uniformly over all patches instead of blockwise masking [50]. Proper selection of data augmentation methods used to create global and local views is critical for generating salient image representations in SSL [51]. Thus, we took particular care to select data augmentations for 3D adaptation.

Since the pretraining dataset includes non-quantitative imaging modalities, image normalization is conducted by linearly mapping the $0.05^{th}$ and $99.95^{th}$ percentiles of intensity to -1 and 1 respectively. Random image augmentations used for pretraining on RGB images include flipping, blurring, converting to grayscale, solarization, and color jitter. Since the color-related augmentations cannot be implemented for 1-channel inputs, we instead utilize medical imaging-related augmentations that may produce robust representations to domain shift including random contrast adjustment, additive noise, gibbs noise, and histogram shift.

The RandomResizedCrop augmentation used in the original DINOv2 randomly crops a portion of an image and resizes it to a specified size, keeping the crop within a range of aspect ratios (ensuring that a crop is not too long or wide). This could not be directly adapted to 3D, as the images that form our pretraining dataset have highly variable slice thicknesses and voxel sizes. For example, the Healthy-Total-Body-CTs [52,53] dataset contains on the order of ~1000 CT slices across a single image. On the other hand, the NYU fastMRI knee [40,41] dataset contains ~30 slices per image. Maintaining a reasonable aspect ratio between the in-plane image dimensions and the out-of-plane (depth) image dimension would be difficult for both datasets simultaneously. Thus, rather than enforcing approximately isotropic spacing and an aspect ratio near one on a randomly cropped and resized volume, we crop the two in-plane image dimensions using the standard 2D RandomResizedCrop, and independently sample the cross-slice dimension crop size. Though this may mean that the cropped volume resulting from this data augmentation can be stretched or squashed in the out-of-plane axis, we expect that using a large variety of pretraining datasets will allow the model to learn to generalize to various volume sizes. This formulation additionally preserves the “local-to-global” correspondences that were relevant in the original DINO [24], where global views take up a larger portion of the original volume than local views.

Finally, we ensured that 3D adaptation code was written with minimal adjustments to low-level modules to integrate and take advantage of all efficiency improvements introduced by the original DINOv2. We term the final model obtained after 3DINO pretraining the 3DINO-ViT.

## 3DINO pretraining implementation details

Our implementation uses PyTorch (https://pytorch.org/) and builds on the GitHub repository released by the original DINOv2 authors (https://github.com/facebookresearch/dinov2). We use MONAI (https://monai.io/) for data loading and implementations of data augmentations. SSL pretraining was conducted on four A100-SXM4-80GB GPUs. All experiments used a ViT-Large [29] and a patch size of 16×16×16. Standard pretraining experiments used a batch size per GPU of 128 (512 total), a global crop size of 96×96×96, a local crop size of 48×48×48, and a base learning rate of 0.002. The EMA parameter $\lambda$ is increased from 0.992 to 1.000 in a cosine schedule. Pretraining progresses for 125,000 iterations over approximately nine days.

We implemented the high-resolution adaptation stage as per the recommendations of the original work, by keeping parameter scheduling the same as pretraining but compressed to progress over 12,500 training iterations instead of the original 125,000 iterations. High-resolution adaptation used a batch size per GPU of 64 (256 total), a global crop size of 112×112×112, a local crop size of 64×64×64, and a base learning rate of 0.001. Adaptation began from the weights learned in the 112,500$^{th}$ pretraining iteration and took approximately two days. Additional hyperparameters can be found in **Suppl. Table 25-26**.

## SOTA comparisons

Tang et al. [21] proposed a SSL pretraining method for 3D medical images, specifically for CT data, using a Swin Transformer [54] backbone. Their method was trained on 5,050 publicly available CT images. We use their publicly released pretrained weights as one baseline comparison against our proposed pretraining method ('Swin Transfer'), and take a randomly initialized Swin Transformer ('Swin Random') to determine the relative benefit of using their pretrained weights.

However, since the Swin Transfer network differs from 3DINO-ViT in both model architecture and pretraining dataset, we take the Swin pretraining implementation, and reimplement it for a vanilla ViT. We then use the reimplemented method to pretrain a vanilla ViT on the 3DINO-ViT pretraining dataset. This forms a separate comparison that specifically investigates the difference in quality of pretraining algorithms ('MONAI-ViT'). The original Swin ViT pretraining method uses the inpainting, contrastive coding, and rotational prediction tasks jointly. Similarly to the original method, the inpainting task is performed by upsampling the ViT patch tokens using transposed convolutions and comparing the reconstructed output to the original image via an L1 loss. The contrastive and rotational tasks are image-level tasks, hence we pass the output of the ViT $[\mathrm{CLS}]$ token to the contrastive coding and rotation prediction linear heads. By doing so, we are also able to more explicitly train an image-level representation for classification experiments.

## MONAI-ViT pretraining implementation details

The Swin ViT pretraining code was taken from the original work [3]. We minimally adjusted the data loading code provided, and only added additional transforms to change intensity scaling to the percentile-based method used for 3DINO-ViT. This was done because the original work was only pretrained on CT images, and thus the scaling range in Hounsfield Units (HU) could be fixed between images. When adding additional non-quantitative scans like relaxation time-weighted MRI sequences to the dataset, the intensity scaling method must also be adjusted. To create a fair comparison, SSL pretraining was also conducted on four A100-SXM4-80GB GPUs. We pretrained a ViT-Large on the same pretraining dataset that was used to train 3DINO-ViT. The method was pretrained using a patch size of 16×16×16 and an image size of 96×96×96. We used a batch size per GPU of 16 (64 total) and pretraining ran for 100,000 iterations over approximately 10 days.

# Finetuning Methods

## Brain Tumor Segmentation (BraTS) dataset

The BraTS 2021 training dataset [32] is a widely-used MRI brain segmentation benchmark. This dataset consists of 1251 patients, each with four types routinely required MRI scans: T1-weighted, T2-weighted, T1-weighted with gadolinium contrast, and T2 Fluid-attenuated Inversion Recovery (FLAIR). These scans were skull-stripped and coregistered. For each patient, medical experts manually generated pixel-level segmentation labels that were combined into Whole Tumor (WT), Tumor Core (TC) and Enhancing Tumor (ET) regions. To form our finetuning dataset, we first removed scans in the dataset taken from TCGA-GBM [55] or TCGA-LGG [56] (which are present in the pretraining dataset) to avoid unfair bias in evaluation. This resulted in 1084 patients that we randomly split into train (N = 758), validation (N = 108) and test (N = 218) sets.

## Beyond the Cranial Vault (BTCV) dataset

The BTCV dataset is a commonly employed CT abdominal organ segmentation benchmark[33]. This dataset consists of abdominal CT scans taken from 30 healthy patients with manual labels generated of 13 organs. These were randomly split into train (N = 20), validation (N = 4) and test (N = 6) sets.

## International Consortium for Brain Mapping (ICBM) dataset

We employed the publicly released ICBM dataset as an MRI brain age classification benchmark [36]. This dataset consists of T1-weighted brain MRI scans of 639 healthy patients with 1339 scans from a variety of ages between 18 and 80. To maintain a reasonably balanced dataset, we binned data into four bins of width 10 between 20 and 60 years of age, discarding scans that did not fall into this range. We split randomly on the patient level to obtain train (N=756), validation (N=151) and test (N=233) scans. Only skull-stripping was performed for data preprocessing using the iCVMapp3r pipeline [57]. For the four bins: [20, 30), [30, 40), [40, 50), [50, 60], the train set contains 335, 199, 108, 144 volumes, the validation set contains 63, 25, 39, 24 volumes, and the test set contains 110, 49, 21, 53 volumes respectively.

## COVID-CT-MD dataset

We used the COVID-CT-MD dataset as a lung CT classification benchmark between COVID-19, Community Acquired Pneumonia (CAP), and healthy patients [37]. This dataset consists of 305 patients with one lung CT scan each that we split randomly into train (N=214), validation (N=30) and test (N=61) scans. For the three classes, Healthy, COVID-19, CAP, the train set contains 54, 121, 39 scans, the validation set contains 7, 15, 8, and the test set contains 15, 33, 13 scans respectively.

## Left Atrium Segmentation Challenge (LA-SEG) dataset

We took the LA-SEG challenge dataset as a left atrium MRI segmentation benchmark [34]. The heart makes up a very small subset of the pretraining dataset, hence this data is used to evaluate the generalizability of the method to an out-of-domain organ. This dataset consists of 154 heart MRI scans from 60 patients and segmented for the left atrial cavity. The challenge dataset was originally split on the patient level-into training (N=100) and test (N=54) sets. We randomly split the training dataset further into subsets used to train (N=80) and validate (N=20) finetuning networks.

## Tumor Detection, Segmentation and Classification Challenge on Automated 3D Breast Ultrasound (TDSC-ABUS) dataset

We used the TDSC-ABUS training dataset as a 3D breast US lesion segmentation benchmark [35]. Ultrasound images have a very different appearance to MRI and CT images and were not present in the pretraining dataset at all. Hence, we use this data to evaluate the generalizability of the method to a completely out-of-domain downstream task. The dataset consists of 100 breast US scans from an unreleased number of patients, with expert segmentations for lesions. We split the data randomly into train (N=70), validation (N=10), and test (N=20) sets.

## 3D ViT-Adapter

ViTs are typically more difficult to train relative to convolutional neural networks (CNNs), especially in a supervised setting with limited training data. This has been attributed to the lack of inductive biases in ViTs and their larger number of trainable parameters [58]. In 3D medical imaging, this has been partially overcome using vision-specific transformer networks, such as the SwinUNETR [21], which currently represents a state-of-the-art (SOTA) in image segmentation. However, the formulation of vanilla ViTs enables many of the improvements introduced in DINOv2 (such as the patch-level objective, their version of FlashAttention, and sequence packing [28]), and has been shown to scale well with dataset sizes [59]. Hence, instead of using a vision-specific network, we convert the ViT-Adapter [30,59] – a popular pretraining-free module that injects spatial information into standard ViT networks – to 3D medical imaging inputs.

The ViT-Adapter was originally proposed for 2D pretrained ViT networks as a way to introduce image-based inductive biases into the network. By building on top of a vanilla ViT, the method is able to take advantage of large-scale pretraining methods [30]. This method uses a simple convolutional network called a Spatial Prior Module to extract local multi-scale spatially relevant features from the original input. It uses a Spatial Feature Injector (‘Injector’) to introduce the extracted multi-scale spatial features into the features obtained from the pretrained ViT. A Multi-Scale Feature Extractor (‘Extractor’) is then used to adapt the multi-scale features based on the pretrained ViT features. Importantly, by using multi-scale features, the method is able to output a feature pyramid much like typical convolutional encoder networks [60]. Overall, the ViT-Adapter is able to greatly improve vanilla ViTs for dense segmentation tasks and was used in the original DINOv2 work as well.

To convert this method to 3D inputs, we first adjusted the Spatial Prior Module to use 3D convolutions for extracting multi-scale features in 3D. To avoid resampling errors with the input sizes of the pretrained ViT network (112×112×112), instead of using feature maps with 1/8, 1/16 and 1/32 of the original spatial input size (as 32 does not divide 112 evenly), we instead used the scales 1/4, 1/8 and 1/16. Thus, the output of the spatial prior module is $\mathcal{F}_{sp} \in \mathbb{R}^{(\frac{HWD}{4^3}+\frac{HWD}{8^3}+\frac{HWD}{16^3})\times F}$, for height, $H$, width, $W$, and depth, $D$, of the input volume and the Transformer feature size, $F$.

The Injector and Extractor networks both rely on the Multi-Scale Deformable Attention (MSDA) formulation of sparse attention [31]. Instead of having both query and keys in standard self-attention enumerate all possible spatial locations in an input image, each query in MSDA only attends to a fixed, small number of keys ( $K = 4$ ). Then, the value features are obtained by sampling the feature map at learnable offset locations. We converted MSDA to 3D inputs by inputting 3D feature maps, and learning an additional deformable offset for the depth axis. We adapted the core 3D deformable attention operation to permit 3D inputs, and enabled non-integer deformable offsets by performing trilinear interpolation on 3D feature maps.

Our implementation makes key changes from the original MSDA when initializing bias parameters for the linear projection predicting the deformable offset. The original 2D work offsets bias for each attention head so that the initial offsets have equal angular separation. For example, with 8 attention heads, the initial offset per head, $O_{init-2D}$, is:

$$O_{init-2D} = \{(-k,-k),(-k,0),(-k,k),(0,-k),(0,k),(k,-k),(k,0),(k,k)\}$$

For each key, $k \in \{1, 2, ..., K\}$, making the angular separation for each offset vector 45 degrees. The key intention of this form of initialization is to ensure more even coverage of the feature map when sampling for value features. With no direct way to extend this into 3D without introducing an intractable number of attention heads, we fix the attention heads to 8, and initialize the bias to have initial offsets pointing towards each 3D *octant*. Concretely, the initial offset per head in 3D MSDA, $O_{init-3D}$, is:

$$O_{init-3D} = \{(-k,-k,-k),(-k,-k,k),(-k,k,-k),(-k,k,k),(k,-k,-k),(k,-k,k),(k,k,-k),(k,k,k)\}$$

Per key, $k$. The difference between 2D and 3D initialization is visually demonstrated in **Extended data Fig. 7**.

## Multi-channel inputs

Pretraining experiments were conducted on single-channel input images. During pretraining, multi-channel unlabelled inputs were split into separate input images. However, some downstream tasks in medical imaging (such as BraTS) benefit from using multiple co-registered modalities to provide complementary information and contrast. The issue of adapting a single-channel pretrained network to multi-channel inputs rarely arises for large 2D natural image datasets, as pretraining and all downstream tasks tend to remain in RGB color space (3 channels).

We adopted a simple channel mixing method to address this. To make full use of the pretrained weights, which have been specifically tuned on single-channel inputs, we did not adjust the patch embedding layer of the ViT. Instead, we passed each image channel through the network individually and obtained a feature vector per channel. These features were then concatenated and passed through a linear layer mapping back to the original transformer feature size, and a Gaussian error linear unit (GELU) [61] activation function. The resulting patch-level feature vectors are passed to the decoder network for downstream dense segmentation tasks.

In practice, we can parallelize the operation passing each channel through the network independently by reshaping the input so that channels form part of the mini-batch (i.e. an input of shape $\mathbb{R}^{B \times C \times H \times W \times D}$ is reshaped to $\mathbb{R}^{(BC) \times 1 \times H \times W \times D}$ for batch size, $B$ ). Generally, we expect the spatial information of multi-channel inputs to be relatively similar between co-registered channels. Thus, to maintain tractability and reduce redundancy when training the ViT-Adapter, the multi-channel features output by the frozen pretrained Transformer blocks were averaged along the channel dimension before being passed to the Injector and Extractor modules. Then, the resulting spatial information output from the Injector is copied along the channel dimension before being added to the Transformer features (**Extended data Fig. 8**). After being passed fully through the ViT, these features were also concatenated and linearly mapped to the original Transformer feature size. As expected, this led to marked benefits even when we used single-channel pretrained weights on multi-channel segmentation tasks.

## 3DINO segmentation

3DINO segmentation experiments were conducted using the 3DINO-ViT weights learned from high-resolution adaptation. 3DINO-ViT was frozen, and the ViT-Adapter module and a UNet-like convolutional decoder were trained on the dense segmentation task (**Extended data Fig. 1a**). Corresponding to the pretraining input size, these experiments also used inputs of size 112×112×112. To evaluate the label efficiency of the pretraining method, we extracted a random subset of the finetuning train sets (with the same random subset taken between experiments).

These finetuning experiments used a batch size of 8, and were conducted on one A100-SXM4-80GB GPU. The base learning rate was set to 0.0001, and finetuning was conducted for 30,000 iterations (regardless of input dataset size). We used the AdamW [62] optimizer with default β and weight decay and a LinearWarmupCosineAnnealing scheduler with 3,000 warmup iterations. For BraTS, we used the Dice loss function, and for BTCV, LA-SEG, and TDSC-ABUS, we used Dice-Cross-Entropy. For all finetuning experiments, we used the validation set to select the best model epoch from training and reported results on the test set. To create final segmentation logits for testing, we strided a sliding window over the full image with an overlap between images of 0.75.

The decoder took in the four-level feature pyramid output from ViT-Adapter and used UNet-like transposed convolutions for upsampling followed by encoder-decoder connections [60]. The decoder consisted of four layers with feature size 256, 128, 64, 32 before mapping to the number of segmentation classes. The ViT-Adapter broke the 3DINO-ViT encoding layers into four blocks each containing 6 Transformer layers, and used 8 MSDA heads. The feature size for ViT-adapter operations was 256, or 25% of the full ViT feature size to reduce computational complexity.

## SOTA segmentation comparisons

The Random encoder network is converted to perform segmentation by adding a convolutional decoder taken from Hatamizadeh et al. [63] and adapted to a ViT-Large by taking the output of the 6th, 12th, 18th, and 24th ViT layer (**Extended data Fig. 1d**). Both the encoder and decoder were tuned end-to-end. These networks otherwise used the same parameters as pretrained initialization.

Segmentation experiments using the SOTA Swin Transfer and Swin Random encoders are conducted by attaching the SwinUNETR decoder network [21] (**Extended data Fig. 1b, 1e**). As done in the original work, the full SwinUNETR was tuned end-to-end. The input size used for segmentation corresponded to the image size used for pretraining, 96×96×96. The same subsets of the finetuning datasets used for 3DINO segmentation were taken in these experiments. Experiments tuning the Swin Transfer and Swin Random networks on BTCV used a batch size of 8, with BraTS experiments using a batch size of 4 (largest power of 2 without running into memory errors). All experiments are conducted on one A100-SXM4-80GB GPU. Since the original network was trained on single-channel inputs, we employed the same multi-channel adaptation strategy for experiments on BraTS. The pretrained weights were originally trained on CT images with intensity normalized to a range of [0, 1]. Segmentation experiments using the Swin Transfer network are thus also normalized to this range to better take advantage of pretraining. All other parameters remained consistent with 3DINO segmentation experiments.

We performed segmentation using the MONAI-ViT network in the same way as the 3DINO-pretrained network (**Extended data Fig. 1c**). The pretrained ViT was frozen, with the ViT-Adapter and decoder being trained. The input size for these experiments was 96×96×96 to match pretraining image size. All other parameters were consistent.

## Linear decoder segmentation

To perform the lightweight linear decoder experiments for segmentation, the pretrained network was frozen, and a two-layer linear network was trained. The first linear layer was the multi-channel projection linear layer. The second layer mapped to the number of segmentation classes, taking concatenated patch representations from the final four ViT layers as input. The output of the linear decoder was a low-resolution volume of class logits (for example, for an input image of size 96×96×96 and a patch size of 16×16×16, the model would output a 6×6×6 map). This volume was upsampled using trilinear interpolation to the original image size to obtain pixel-wise segmentation logits, which are compared with the ground truth mask (**Extended data Fig. 9**). The linear decoder experiments used a base learning rate of 0.001 and a batch size of 16. Other parameters, including optimizer, scheduler, iterations trained, and loss functions remained consistent with 3DINO segmentation experiments. We did not use Swin Transfer and Swin Random for these experiments as their final representation map was highly downsampled (only a 3×3×3 map for the 96×96×96 input).

## Linear classification probing

Classification experiments are conducted with the three pretrained models investigated in this study: 3DINO-ViT, MONAI-ViT, and Swin Transfer. In all cases, linear probing on frozen pretrained weights was performed similarly to the original DINOv2 by using a grid search on three key parameters: the learning rate, the number of final ViT layer outputs to concatenate, and whether the averaged patch tokens are concatenated to the $[\text{CLS}]$ token. As in the original work, the learning rates were searched in the set $\{0.0001, 0.0002, 0.0005, 0.001, 0.002, 0.005, 0.01, 0.02, 0.05, 0.1, 0.2, 0.3, 0.5\}$, number of output layers in $\{1, 4\}$ and averaged patch token concatenation in $\{True, False\}$. The best parameters based on validation performance were then used for evaluating on the test set.

The Swin Transfer network does not train a $[\text{CLS}]$ token for forming image-level representations. To probe the image-level representations from these pretrained weights, we extracted the output from the pretrained contrastive coding head of the network. As with segmentation experiments, the input images to the pretrained Swin Transformer network were normalized between [0, 1].

Classification experiments were conducted for 12,500 iterations with input sizes that match what was originally used to pretrain the models. All experiments used a batch size of 32, and were conducted on one A100-SXM4-80GB GPU. We used the SGD optimizer with a momentum of 0.9 and 0 weight decay, a cosine annealing learning rate scheduler, and Cross-Entropy loss. As with segmentation experiments, we extracted a random subset of the finetuning train sets to test reliance on labeled data.

# Visualization Methods

No images were registered to atlases for pretraining, finetuning, finetuning visualizations, or unsupervised visualizations.

## Principal component analysis (PCA)

To create our PCA visualizations, we took the features output by 3DINO-ViT at each 16×16×16 patch location. We flattened the spatial dimension of these features and performed PCA to reduce the feature dimension. The first PCA component typically separates the image foreground and background, and a simple threshold was taken to color background regions for visualization. Then, the next three PCA components at each patch location were represented by the red, green, and blue channels of an RGB image, respectively. This color-coded representation offers an intuitive way to interpret the contribution of each principal component (variance axis) to the overall structure of the images.

## Multi-head self-attention (MHSA)

We extracted the MHSA map of one attention head on the $[\text{CLS}]$ token of the final 3DINO-ViT Transformer layer to visualize regions of interest. This map describes the relative attention given to each patch for generating image-level representations.

## Statistical analysis

Error bars and 95% confidence intervals are computed using 1000 bootstrapped samples from scan-wise Dice and HD95 metrics (for segmentation) and AUC and F1 scores from five independent and randomized runs (for classification). Subsets drawn from the full labeled dataset are randomized across all five classification runs, but remain consistent across the compared pretraining methods. For segmentation results, statistical significance is computed using a paired nonparametric Wilcoxon test on Dice score and HD95 obtained per scan. For classification, significance is computed via an unpaired Welch's t-test against AUC and F1 scores from the five runs. Statistical tests are implemented in relevant Python libraries (Scipy, Seaborn), and visualizations are created using the Statsannotations library.

## Acknowledgements

This work was supported by funding from the Natural Sciences and Engineering Research Council (NSERC) Discovery grant (RGPIN-2021-03728), Canada Foundation for Innovation (40206), and the Ontario Research Fund. T.X. is funded by the NSERC PGS-D award. M.G. is funded by the Canada Research Chairs program award (CRC-2021-00374). A.L.M. is supported by the Tory Family Chair in Oncology. We are grateful to the Digital Research Alliance of Canada (alliancecan.ca) for their allocation of computing resources used in parts of this research. We are grateful for support from the Black Centre for Brain Resilience and Recovery and the Harquail Centre for Neuromodulation.

A large portion of the pretraining dataset was obtained from The Cancer Imaging Archive [64].

Data used in the preparation of this article were obtained from the Alzheimer's Disease Neuroimaging Initiative (ADNI) database (adni.loni.usc.edu). The ADNI was launched in 2003 as a public-private partnership, led by Principal Investigator Michael W. Weiner, MD. The primary goal of ADNI has been to test whether serial magnetic resonance imaging (MRI), positron emission tomography (PET), other biological markers, and clinical and neuropsychological assessment can be combined to measure the progression of mild cognitive impairment (MCI) and early Alzheimer's disease (AD).

Data collection and sharing for the Alzheimer's Disease Neuroimaging Initiative (ADNI) is funded by the National Institute on Aging (National Institutes of Health Grant U19 AG024904). The grantee organization is the Northern California Institute for Research and Education. In the past, ADNI has also received funding from the National Institute of Biomedical Imaging and Bioengineering, the Canadian Institutes of Health Research, and private sector contributions through the Foundation for the National Institutes of Health (FNIH) including generous contributions from the following: AbbVie, Alzheimer's Association; Alzheimer's Drug Discovery Foundation; Araclon Biotech; BioClinica, Inc.; Biogen; Bristol-Myers Squibb Company; CereSpir, Inc.; Cogstate; Eisai Inc.; Elan Pharmaceuticals, Inc.; Eli Lilly and Company; EuroImmun; F. Hoffmann-La Roche Ltd and its affiliated company Genentech, Inc.; Fujirebio; GE Healthcare; IXICO Ltd.; Janssen Alzheimer Immunotherapy Research & Development, LLC.; Johnson & Johnson Pharmaceutical Research & Development LLC.; Lumosity; Lundbeck; Merck & Co., Inc.; Meso Scale Diagnostics, LLC.; NeuroRx Research; Neurotrack Technologies; Novartis Pharmaceuticals

Corporation; Pfizer Inc.; Piramal Imaging; Servier; Takeda Pharmaceutical Company; and Transition Therapeutics.

Data were provided (in part) by the Human Connectome Project, WU-Minn Consortium (Principal Investigators: David Van Essen and Kamil Ugurbil; 1U54MH091657) funded by the 16 NIH Institutes and Centers that support the NIH Blueprint for Neuroscience Research; and by the McDonnell Center for Systems Neuroscience at Washington University.

We would like to acknowledge the individuals and organizations that have made data used for this research available including, the Ontario Brain Institute, the Brain-CODE platform, the Government of Ontario. Matching funds were provided by participant hospital and research foundations, including the Baycrest Foundation, Bruyere Research Institute, Centre for Addiction and Mental Health Foundation, London Health Sciences Foundation, McMaster University Faculty of Health Sciences, Ottawa Brain and Mind Research Institute, Queen's University Faculty of Health Sciences, the Thunder Bay Regional Health Sciences Centre, the University of Ottawa Faculty of Medicine, University Health Network, Sunnybrook, and the Windsor/Essex County ALS Association. The Temerty Family Foundation provided the major infrastructure matching funds.

Data used in the preparation of this work were obtained (in part) from the International Consortium for Brain Mapping (ICBM) database (www.loni.usc.edu/ICBM). The ICBM project (Principal Investigator John Mazziotta, M.D., University of California, Los Angeles) is supported by the National Institute of Biomedical Imaging and BioEngineering. ICBM is the result of efforts of co-investigators from UCLA, Montreal Neurologic Institute, University of Texas at San Antonio, and the Institute of Medicine, Juelich/Heinrich Heine University - Germany.

Data were provided (in part) by OASIS-3: Longitudinal Multimodal Neuroimaging: Principal Investigators: T. Benzinger, D. Marcus, J. Morris; NIH P30 AG066444, P50 AG00561, P30 NS09857781, P01 AG026276, P01 AG003991, R01 AG043434, UL1 TR000448, R01 EB009352. AV-45 doses were provided by Avid Radiopharmaceuticals, a wholly owned subsidiary of Eli Lilly.

Data used in the preparation of this article were obtained (in part) from the NYU fastMRI Initiative database [40,41]. As such, NYU fastMRI investigators provided data but did not participate in analysis or writing of this report. A listing of NYU fastMRI investigators, subject to updates, can be found at fastmri.med.nyu.edu. The primary goal of fastMRI is to test whether machine learning can aid in the reconstruction of medical images.

## Data availability

All external data used in this study can be obtained online. DOIs and links to pretraining datasets can be found in **Extended data Table 1**. Finetuning datasets can be found as follows: BraTS (https://www.synapse.org/Synapse:syn25829067/wiki/610863), BTCV (https://www.synapse.org/Synapse:syn3193805/wiki/89480), LA-SEG (https://www.cardiacatlas.org/atriaseg2018-challenge/atria-seg-data/), TDSC-ABUS (https://tdsc-abus2023.grand-challenge.org/), ICBM (https://ida.loni.usc.edu/collaboration/access/appLicense.jsp), COVID-CT-MD

(https://figshare.com/collections/COVID-CT-MD_COVID-19_Computed_Tomography_CT_Scan_Dataset_Applicable_in_Machine_Learning_and_Deep_Learning/5129081).

## Code availability

Full 3DINO code can be found at https://github.com/AICONSlab/3DINO with documented instructions for performing pretraining, finetuning, unsupervised visualization, and simple model inference. 3DINO-ViT model weights will be made available upon paper acceptance.